
\magnification=\magstep1
\parindent=15pt
\centerline{\bf DIMENSIONAL EXPANSION FOR THE ISING LIMIT}
\centerline{\bf OF QUANTUM FIELD THEORY}
\bigskip
\bigskip
\bigskip
\centerline{Carl M. Bender and Stefan Boettcher\footnote{$^*$}{Current
Address: Brookhaven National Laboratory; Upton, NY 11973}}
\medskip
\centerline{Department of Physics}
\medskip
\centerline{Washington University}
\medskip
\centerline{St. Louis, MO 63130}
\bigskip
\bigskip
\bigskip
\bigskip
\bigskip
\centerline{\bf ABSTRACT}
\bigskip

A recently-proposed technique, called the dimensional expansion, uses the
space-time dimension $D$ as an expansion parameter to extract nonperturbative
results in quantum field theory. Here we apply dimensional-expansion
methods to examine the Ising limit of a self-interacting scalar field theory.
We
compute the first few coefficients in the dimensional expansion for
$\gamma_{2n}$, the renormalized $2n$-point Green's function at zero momentum,
for $n\!=\!2$, 3, 4, and 5. Because the exact results for
$\gamma_{2n}$ are known at $D\!=\!1$ we can compare the predictions of the
dimensional expansion at this value of $D$. We find typical errors of less
than $5\%$. The radius of convergence of the dimensional expansion for
$\gamma_{2n}$ appears to be ${{2n}\over {n-1}}$. As a function of the
space-time dimension $D$, $\gamma_{2n}$ appears to rise monotonically with
increasing $D$ and we conjecture that it becomes infinite at
$D\!=\!{{2n}\over {n-1}}$.  We presume that for values of $D$ greater than this
critical value, $\gamma_{2n}$ vanishes identically because the corresponding
$\phi^{2n}$ scalar quantum field theory is free for $D\!>\!{{2n}\over{n-1}}$.
\footnote{}{PACS numbers: 11.10.-z, 11.90.+t, 02.90.+p}
\footnote{}{hep-th/9311060}
\vfill \eject

In a recent letter[1] we
proposed a new technique called the dimensional expansion, which can be used
to obtain nonperturbative results in quantum field theory. The dimensional
series uses the space-time dimension $D$ as an expansion parameter. The first
term in such an expansion is easy to obtain because a quantum field theory can
be solved in closed form in zero-dimensional space-time. An advantage of
dimensional expansions is that some of the nontrivial aspects of the
interaction
already
appear at $D\!=\!0$. (Traditional perturbative methods yield only
noninteractive
results in leading order.) The obvious question is how one can obtain the
coefficients of higher powers of $D$. A detailed explanation of how to do so
is given in a subsequent paper[2].

Here we use the dimensional expansion to compute the first four $\gamma_{2n}$,
the renormalized $2n$-point Green's functions at zero external momentum, for a
self-interacting scalar quantum field theory in the Ising limit. Specifically,
we calculate $\gamma_4$ to fourth order in powers of $D$, $\gamma_6$ to fifth
order in powers of $D$, $\gamma_8$ to sixth order in powers of $D$, and
$\gamma_{10}$ to seventh order in powers of $D$:
$$
\eqalign{
\gamma_4~&=~{1\over 12} \bigl[ 1+(1.180\pm0.001) D +(0.620\pm0.001) D^2 +
   (0.18\pm0.02) D^3 \cr
&~~~~~~~~~~~~+ (0.03\pm 0.02) D^4 +\ldots \bigr ]~~,\cr
\gamma_6~&=~{1\over 30} \bigl[ 1+(2.20\pm0.02) D + (2.30\pm0.03) D^2 +
   (1.50\pm0.03) D^3 + (0.55\pm0.04) D^4 \cr
&~~~~~~~~~~~~+ (0.12\pm0.04) D^5 +\ldots \bigr ] ~~,\cr
\gamma_8~&=~{1\over 56} \bigl[ 1+(3.0\pm0.1) D + (4.5\pm0.1) D^2 + (4.2\pm0.1)
D^3 + (2.6\pm0.1) D^4 \cr
&~~~~~~~~~~~+ (1.2\pm0.2) D^5 + (0.6\pm0.2) D^6 +\ldots \bigr ] ~~,\cr
\gamma_{10}~&=~{1\over 90} \bigl[ 1+(4.11\pm0.02)D + (8.0\pm0.1) D^2 +
(10.0\pm0.3) D^3 + (8.0\pm0.3) D^4 \cr
&~~~~~~~~~~~+ (4.5\pm0.3) D^5 + (1.8\pm0.3) D^6+(0.7\pm0.4)D^7 +\ldots  \bigr ]
{}~. \cr} \eqno(1)
$$

To obtain these dimensional
expansions we use the graphical methods described in Ref.~2. These graphical
methods rely on lattice strong-coupling techniques that were developed and
explained in an earlier series of papers[3,4,5,6]. For the Lagrangian
$$
{\cal L}~=~{1 \over 2} [{\partial \phi(x)}]^2
          ~+~{1 \over 2} m^2 {\phi(x)}^2 ~+~{1\over 4}g {\phi(x)}^4\eqno(2)
$$
the Ising limit[7,8,9] is defined as the limit in which the unrenormalized
coupling constant $g$ tends to infinity while the renormalized mass, $M$, is
held
fixed. The Ising limit is conveniently obtained by choosing $m^2\!\propto\!-g$.
In the limit $g\!\to\!\infty$ the theory asymptotically approaches a two-state
system. The Green's functions of this system are universal in the sense that
they are independent of the power of $\phi$ in the self-interaction term in
(2); $g\phi^{2k}$ gives the same results as $g\phi^4$ for all $k\!\geq\!2$.

Lattice strong-coupling methods are especially well suited to obtain the
dimensional expansion of Green's functions in quantum field theory because the
lattice integral for each graph is a {\it polynomial} in powers of $D$. This
property leads to an efficient organization of the graphs that contribute to
each order in the $D$-series. We achieve a high order in the graphical
expansion
by eliminating all graphs except those that contribute to the coefficients in
the dimensional expansion under consideration. We employ an intermediate
renormalization scheme to calculate the renormalized mass $M$ and dimensionless
renormalized $2n$-point scattering amplitudes $\gamma_{2n}$ at zero momentum.
We perform mass renormalization of the scattering amplitudes by eliminating the
bare mass $m$ in $\gamma_{2n}$ in favor of the renormalized mass $M$. We then
use Pad\'e extrapolation methods to derive a sequence of approximants for each
coefficient in the dimensional expansions in (1) for each of the scattering
amplitudes $\gamma_{2n}$ in the continuum limit. We believe that each
Pad\'e sequence gives an accurate approximation to the true
coefficient in the dimensional expansion for $\gamma_{2n}$ because these
dimensional series are in good numerical agreement with known exact results for
$\gamma_{2n}$[3]. The series in (1) are exact at $D\!=\!0$. At $D\!=\!1$ the
exact results are $\gamma_4\!=\!{1\over4}$, $\gamma_6\!=\!{1\over4}$,
$\gamma_8\!=\!{5\over{16}}$, $\gamma_{10}\!=\!{7\over{16}}$, and the results
for
(1) are $\gamma_4=0.250\pm0.003\,(1\%$ error),
$\gamma_6=0.26\pm0.01\,(4\%$ error), $\gamma_8=0.31\pm0.01\,(4\%$ error), and
$\gamma_{10}=0.42\pm0.02\,(5\%$ error).

We obtain the graphical rules for the lattice strong-coupling expansion by
observing that in the limit of large $g$ the kinetic term in the Lagrangian (2)
can be viewed as a small perturbation. Therefore, the generating function
$$
{\cal Z}[J]~=~{\cal N}~\int{\cal D} \phi(x)~exp \Bigl\{ ~-\textstyle \int\!
 d^{ D}x \bigl[ {1 \over 2} [{\partial \phi(x)}]^2
                    ~+~{1 \over 2} m^2 {\phi(x)}^2 ~+~{1\over 4}g {\phi(x)}^4
                    ~-~J(x) {\phi(x)} \bigr] \Bigr\} \eqno(3)
$$
for the quantum field theory associated with the Lagrangian (2)
can be rewritten as
$$
{\cal Z}[J]~=~exp \Bigl\{ {1 \over 2} \textstyle \int\!
d^{ D}xd^{ D}y {\textstyle {\delta \over {\delta J(x)}}}
                {\cal D}^{-1}(x-y)
{\textstyle {\delta \over {\delta J(y)}}} \Bigr\}~~
                {\cal Z}_0[J]~~, \eqno(4)
$$
where ${\cal D}^{-1}(x-y)={\partial}^2 {\delta}^{ D}(x-y)$ and
$$
{\cal Z}_0[J]~=~{\cal N} \int\! {\cal D} \phi~
                  exp \Bigl\{-\textstyle \int\! d^{ D}x \bigl[
                  {1 \over 2} m^2 \phi(x)^2
               ~+~{1\over 4}g\phi(x)^4~-~J(x)\phi(x)
                  \bigr] \Bigr\}~~.  \eqno(5)
$$

The factorization in (4) of the partition function leads to the
strong-coupling lattice expansion. By introducing a $D$-dimensional hypercubic
lattice with lattice spacing $a$ we  rewrite (5) as

$${\cal Z}_0[J]~=~{\cal N} \prod_i \int\limits_{-\infty}^{\infty}\! dt~
exp \Bigl\{ - {1 \over 2} a^{ D} m^2 t^2 - {1 \over 4} a^{ D} g t^4
+ a^{ D} J_i t \Bigr\}~~.   \eqno(6)
$$
Next, we expand in powers of $J_i$ and, to obtain the Ising limit, we set
$$
m^2~=~-\alpha g a^{2-D}~~, \eqno(7)
$$
where $\alpha$ is a dimensionless parameter considered to be small in the
strong-coupling limit:
$$
{\cal Z}_0[J]~=~{\cal N} \prod_i \sum_{n=0}^{\infty}
                  {1 \over {(2n)!}} (a^{ D} J_i)^{2n}
                  \int\limits_0^{\infty} dt~t^{n-1/2}~
exp \Bigl\{ - {1 \over 4} a^{ D} g \bigl[ t^2 - 2\,\alpha\,a^{2- D} t
 \bigr]     \Bigr\}~~.  \eqno(8)
$$
In the limit $g\!\to\!\infty$ the integral in (8) is asymptotic to $\alpha^n$
multiplied by a constant independent of $n$ which
we absorb into $\cal N$. Thus, we write ${\cal Z}_0[J]$ in (8) as
$$
{\cal Z}_0[J]~=~{\cal N} exp \Bigl\{ a^{ D} \sum_i \Bigl[
                  \sum\limits_{n=1}^{\infty} {1 \over {(2n)!}}
                  J_i^{2n} V_{2n} \Bigr] \Bigr\}~~,  \eqno(9)
$$
where the vertices are $V_2\!=\!a^2\alpha$, $V_4\!=\!-2a^{4+D}\alpha^2$,
$V_6\!=\!16a^{6+2D}\alpha^3$, $V_8\!=\!-272a^{8+3D}\alpha^4$,
$V_{10}\!=\!7936a^{10+4D}\alpha^5$ and so on. The propagator on the lattice can
be written in vector notation as ${\cal D}^{-1}\!=\!a^{-D-2}[(\underline
1)\!-\!2D(\underline 0)]$. This notation was introduced in Ref.~4 where this
discrete form of
the propagator was used to evaluate lattice integrals. The lattice
strong-coupling expansion is organized by the number of free propagators
${\cal D}^{-1}$ (in contrast to weak-coupling expansions where the number of
{\it vertices} and not the number of lines determines the order).

To compute $\gamma_{2n}$ it is necessary to calculate the
one-particle-irreducible $2n$-point functions
$\Lambda_{2n}$ for $n=1$, 2, 3, 4, and 5, in the strong-coupling
expansion and to find their Fourier transforms $\tilde {\Lambda}_{2n}$ in
momentum space at zero external momentum. We must also compute
${{\partial}\over{\partial (p^2)}} {\tilde \Lambda}_2(p^2)\vert_{p^2=0}$ to
obtain the wave-function renormalization constant defined by
$Z^{-1}\!\equiv\!1\!+\!{\partial \over {\partial (p^2)}}
{\tilde \Lambda}^{-1}_2 \vert_{p^2=0}$.
We define the scattering amplitudes $\gamma_{2n}$ as the {\it dimensionless}
renormalized one-particle-irreducible vertices at zero external momentum
$$
\gamma_{2n}~\equiv~{\tilde \Gamma}^R_{2n}(0,0,\ldots,0)
                M^{D(n-1)-2n} ~~,  \eqno(10)
$$
where $M$ is the renormalized mass defined as
$M^2\!\equiv\!{\tilde \Gamma}^R_2(0,0)$.
There are simple rules giving $\Gamma_{2n}$ in terms of $\Lambda_{2m}$,
$m\leq n$, which are explained in Ref.~4:
$$\eqalign{
  \Gamma_2~&=~\Lambda_2^{-1} ~~,\cr
  \Gamma_4~&=~-~\Lambda_4 \Lambda_2^{-4} ~~,\cr
  \Gamma_6~&=~-~\Lambda_6 \Lambda_2^{-6}
           ~+~{{6!} \over {2~(3!)^2}} \Lambda^2_4 \Lambda_2^{-7}
                                             ~~,\cr
  \Gamma_8~&=~-~\Lambda_8 \Lambda_2^{-8}
           ~+~ {{8!} \over {3!~5!}}
               \Lambda_4 \Lambda_6 \Lambda_2^{-9}
           ~-~{{8!} \over {(2!)^2~(3!)^2}}
               \Lambda^3_4 \Lambda_2^{-10} ~~,\cr
  \Gamma_{10}~&=~-~\Lambda_{10} \Lambda_2^{-10}
              ~+~ {{10!} \over {3!~7!}}
                  \Lambda_8 \Lambda_4 \Lambda_2^{-11}
              ~+~ {{10!} \over {2~(5!)^2}}
                  \Lambda^2_6 \Lambda_2^{-11}
              ~-~ {{10!} \over {2~3!~5!}}
                  \Lambda_6 \Lambda^2_4 \Lambda_2^{-12} \cr
              &~~~~-~ {{10!} \over {2~(3!)^2~4!}}
                  \Lambda_6 \Lambda^2_4 \Lambda_2^{-12}
              ~+~ {{10!} \over {2~(2!)^2~(3!)^2}}
                  \Lambda^4_4 \Lambda_2^{-13}
              ~+~ {{10!} \over {(3!)^4}}
                  \Lambda^4_4 \Lambda_2^{-13}~~. \cr
         } \eqno(11)
$$
We use the wave-function renormalization constant $Z$ to renormalize the
one-particle-irreducible vertices in an intermediate renormalization scheme
according to
${\tilde \Gamma}^R_{2n}(0,\ldots,0)\!=\!Z^n{\tilde \Gamma}_{2n}(0,\ldots,0)$.
In order to mass renormalize the scattering amplitudes $\gamma_{2n}$,
we eliminate the bare mass $m$, which is related to $\alpha$
through (7), in favor of the renormalized mass $M$. To that end, we simply
invert the relation obtained for the renormalized mass
$$\eqalign{
M^2a^2~&=~\alpha^{-1}-2\,D+ (2\,D-{2\over 3} )\alpha+ (4\,D
^{2}-{{26}\over{3}}\,D+{{194}\over{45}} )\alpha^{3} \cr
&~~~~~~~+  (32\,D^{3}-132\,D^{2}+{{2584}\over{15}}\,D-{{68164}\over{945}}
 )\alpha^{5} \cr
&~~~~~~~+ (-2048\,D^{5}-4096\,D^{4}-1024\,D^{3}-
800\,D^{2}+480\,D-64 )\alpha^{6} +\ldots\cr }   \eqno(12)
$$
to expand $\alpha$ in terms of $y\equiv a^{-2}M^{-2}$:
$$\eqalign{
\alpha~&=~y-2\,Dy^2+ (4\,D^2+2\,D-{2\over 3} )y^3
+ (-8\,D^3-12\,D^2+4\,D )y^4 + (16\,D^4+48\,D^3 \cr
&~~~~~~~-4\,D^2-14\,D+{{26}\over5} )y^5 + (-32\,D^5-160\,D^4
    -{{200}\over3}\,D^3+140\,D^2-52\,D ) y^6  \cr
&~~~~~+ (64\,D^6+480\,D^5+560\,D^4-720\,D^3+20\,D^2+272\,D-{{636}\over7}
 )y^7+\ldots~~. \cr}   \eqno(13)
$$
We then substitute (13) for $\alpha$ in every $\Gamma_{2n}^R$ to obtain
$$\eqalignno{
{\gamma_4}~&=~{y^{D/2} \over 12}~\bigr[1+4\,D\,y+ (4\,D^{2}-10\,D
 ) y^{2}+16\,D y^{3}+ (-80\,D^{2}+30\,D )y^{4} + (256\,D^{3} \cr
&~~~~~~~~~~~~~~+104\,D^{2}-192\,D ) y^{5}
+  (-704\,D^{4}-1736\,D^{3}+2508\,D^{2}-656\,D ) y^{6} \cr
&~~~~~~~~~~~~+ (1792\,D^{5}+10432\,D^{4}-11232\,D^{3}-3872\,D^{2}+4992\,D
 ) y^{7} + \ldots \bigr]~~, &(14)\cr
{\gamma_6}~&=~{y^D \over 30}~\bigl[ 1+6\,D\,y+ (12\,D^{2}-6\,D
 ) y^{2}+ (8\,D^{3}-12\,D^{2}-20\,D ) y^{3}\cr
&~~~~~~~~~~~~+ (48\,D^{2}+48\,D ) y^{4} + (-96\,D^{3}-816\,D ^{2}+528\,D
 ) y^{5} \cr
&~~~~~~~~~~~~+ (192\,D^{4} +4640\,D^{3}-2736\,D^ {2}-560\,D ) y^{6} \cr
&~~~~~~~~~~~~+ (-384\,D^{5}-18432\,D^{4}-10800\,D^{3}+46512\,D^{2}-23040\,D )
 y^{7} +\ldots \bigr],&(15) \cr
{\gamma_8}~&=~{y^{3D/2} \over 56}~\bigl[1+8\,D\,y+ (24\,D^{2}-8\,D
 ) y^{2}+ (32\,D^{3}-32\,D^{2} )y^{3} \cr
&~~~~~~~~~~~~+ (16\,D^{4}-32\,D^{3}-36\,D^{2}-18\,D ) y^{4}+ (
896\,D^{2}-448\,D ) y^{5}  \cr
&~~~~~~~~~~~~+ (-4192\,D^{3}-920\,D^{2}+ 2816\,D ) y^{6} \cr
&~~~~~~~~~~~~+ (13184\,D^{4}+52064\,D^{3}-92800\,D^ {2}+38400\,D ) y^{7}
+ \ldots \bigr] ~~, &(16)\cr
{\gamma_{10}}~&=~{y^{2D} \over 90}~\bigl[ 1+10\,D\,y+ (40\,D^{2}-10\,D
 ) y^{2}+ (80\,D^{3}-60\,D^{2} ) y^{3}  \cr
&~~~~~~~~~~~~+  (80\,D^{4}-120\,D^{3}+30\,D ) y^{4}+ (32\,D^{5}
-80\,D^{4}-300\,D^{2}+108\,D ) y^{5}  \cr
&~~~~~~~~~~~~+ (1280\,D^{3}+ 5040\,D^{2}-4240\,D ) y^{6} \cr
&~~~~~~~~~~~~+ (-2560\,D^{4}-64080\,D^{3}+76880\,D^{2}-23040\,D ) y^{7}
+\ldots  \bigr] ~~.&(17)\cr}
$$

The strong-coupling expansions in (14-17) were obtained by treating the
dimensionless parameter $\alpha\!=\!-a^{ D-2} m^2/g$ as small in the limit
where
the bare coupling $g$ tends to infinity. The relation in (12) explicitly
carries the assumption of smallness over to the parameter $y$.
This justifies the reversion of (12) into (13) and the subsequent
reexpansion of the scattering amplitudes $\gamma_{2n}$ in powers
of $y$. Up to this point we have taken the lattice spacing
$a$ to be held fixed. We expect that in the continuum limit $a\!\to\!0$ our
expressions for $\gamma_{2n}$ become the corresponding quantities of the
continuum theory. The continuum limit is subtle because as $a\to 0$ the
parameter $y$ that we have taken to be small actually becomes
infinite.
Hence, subsequent terms in this expansion for the scattering amplitudes
$\gamma_{2n}$ in (14-17) are increasingly singular as a series in powers of $y$
in the limit where $a \to 0$.

We use Pad\'e extrapolation techniques to extract information from
perturbation series like those in (14-17), where the perturbative parameter
tends to infinity. The Pad\'e extrapolation
method employed here uses as input a perturbation series of the form
$$
f(y)~=~y^r~(c_0~+~c_1 y~+~c_2 y^2~+~\ldots~)~~~~~~(r\not=0)~~, \eqno(18)
$$
where we assume that $f(\infty)$ is finite.  We first take the $r$th root of
both sides of (18) and divide by
$y$ to obtain
$$
{f(y)^{1/r} \over y}~=~(new~power~series~in~y)~. \eqno(19)
$$
Then we take the $N$th power of the right side of (19)
for $N=1,2,3,\ldots~$, reexpand and form the $(0,N)$-Pad\'e approximant.
By extracting the coefficient of $y^N$ in the denominator of the
$(0,N)$-Pad\'e and raising it to the power $-r/N$, we create a
sequence of Pad\'e extrapolants for $f(\infty)$[10].
We apply this method to the scattering amplitudes $\gamma_{2n}$ in (14-17) and
expand each $(0,N)$-Pad\'e approximant for all $\gamma_{2n}$ in
powers of $D$. For each $n$, we obtain a sequence in $N$ of
$(0,N)$-Pad\'e approximants for each coefficient in the $D$-series
of $\gamma_{2n}$. In Fig.~1 we plot the $(0,N)$-Pad\'e extrapolants for the
first four coefficients in the dimensional expansion of $\gamma_4$ as functions
of $1/N$.
We indicate the errors in the Pad\'e determination of the coefficients in (1).
We truncate the dimensional expansions for each $\gamma_{2n}$ after that
coefficient for which the estimated error of the following coefficient becomes
significant compared to its absolute size.

Observe that the series in (1) all have positive coefficients and therefore
each
$\gamma_{2n}$ is a monotonically rising function of $D$. Each of these
functions
is plotted in Fig.~2. For each $n$, $\gamma_{2n+2}$ is growing faster than
$\gamma_{2n}$ for increasing $D$. We believe that the radius of convergence of
the $D$-series for $\gamma_{2n}$ is likely to be $D\!=\!{{2n}\over{n-1}}$, the
space-time dimension for which the coupling constant $g$ of a $g\phi^{2n}$
theory becomes dimensionless and the theory becomes renormalizable. Since we
expect that for values of $D\!>{{2n}\over{n-1}}$, $\gamma_{2n}$
vanishes[11,12,13,14,15,16], we assume that there is a singularity (possibly a
natural boundary) in the complex-$D$ plane at $D\!=\!{{2n}\over{n-1}}$.
\vfill \eject

\centerline{\bf REFERENCES}
\bigskip
\item{[1]} C. M. Bender, S. Boettcher, and L. Lipatov, Phys. Rev. Lett.
{\bf 68}, 3674 (1992).
\medskip
\item{[2]} C. M. Bender, S. Boettcher, and L. Lipatov, Phys. Rev. D {\bf 46},
5557 (1992)
\medskip
\item{[3]} C. M. Bender, F. Cooper, G. S. Guralnik, H. Moreno, R. Roskies,
and D. H. Sharp, Phys. Rev. Lett. {\bf 45}, 501 (1980).
\medskip
\item{[4]} C. M. Bender, F. Cooper, G. S. Guralnik, R. Roskies, and D. H.
Sharp,
Phys. Rev. D {\bf 23}, 2976 (1981).
\medskip
\item{[5]} C. M. Bender, F. Cooper, G. S. Guralnik, R. Roskies, and D. H.
Sharp,
Phys. Rev. D {\bf 23}, 2999 (1981).
\medskip
\item{[6]} C. M. Bender, F. Cooper, G. S. Guralnik, R. Roskies, and D. H.
Sharp,
Phys. Rev. D {\bf 24}, 2683 (1981).
\medskip
\item{[7]} G. A. Baker Jr., Phys. Rev. D {\bf 15}, 1552 (1977).
\medskip
\item{[8]} G. A. Baker Jr., in ``Phase Transitions and Critical Phenomena,''
C. Domb and J. L. Lebowitz, eds., Vol. 9, p. 233 (Academic, London, 1984).
\medskip
\item{[9]} G. A. Baker Jr., Phys. Rev. Lett. {\bf 69}, 3264 (1992).
\medskip
\item{[10]} C. M. Bender, Los Alamos Science {\bf 2}, 76 (1981).
\medskip
\item{[11]} M. Aizenman, Phys. Rev. Lett. {\bf 47}, 1 (1981).
\medskip
\item{[12]} M. Aizenman, Commun. Math. Phys. {\bf 86}, 1 (1982).
\medskip
\item{[13]} J. Fr\"ohlich, Nucl. Phys. B {\bf 200} [FS4], 281 (1982).
\medskip
\item{[14]} J. Fr\"ohlich, in ``Progress in Gauge Field Theory,'' eds.
G. 't Hooft, A. Jaffe, H. Lehmann, P. K. Mitter, I. M. Singer and A. Stora,
p. 169 (Plenum, New York, 1984).
\medskip
\item{[15]} M. L\"uscher and P. Weisz, Nucl. Phys. B {\bf 290} [FS20], 25
(1987).
\medskip
\item{[16]} D. Callaway, Physics Reports {\bf 167}, 179 (1988) and
References therein.
\vfill \eject

\centerline{\bf FIGURE CAPTIONS}
\bigskip
\parindent=1in
\item{Figure 1.} Pad\'e extrapolation for the first four
coefficients, $b_i$ ($i=1,\ldots,4)$, in the dimensional expansion of
$\gamma_4\!=\!
{1\over{12}}(1\!+\!b_1D\!+\!b_2D^2\!+\!b_3D^3\!+\!b_4D^4\!+\!\ldots)$.
For each coefficient $b_i$ we plot the value of the $(0,N)$-Pad\'e (shown as
cross) as a function of $1/N$ for $N=1,\ldots,11$. The continuum value of each
$b_i$ is the extrapolation of the sequence to $N\!=\!\infty$. In Eq. (1) we
list
the results of this procedure for $\gamma_4$, $\gamma_6$, $\gamma_8$, and
$\gamma_{10}$.
\bigskip
\item{Figure 2.} Plot of $\gamma_4$, $\gamma_6$, $\gamma_8$, and $\gamma_{10}$
in Eq. (1) as functions of $D$. For each $n$, $\gamma_{2n}$ rises monotonically
and $\gamma_{2n+2}$ is growing faster than $\gamma_{2n}$ for increasing $D$.
We believe that the radius of convergence of the $D$-series for each
$\gamma_{2n}$ in Eq. (1) is $D\!=\!{{2n}\over{n-1}}$.
\bye